\documentclass[a4paper,11pt]{article}
\usepackage{amssymb}
\usepackage{amsfonts}
\usepackage{graphicx}
\usepackage{dcolumn}
\usepackage{amsmath}
\usepackage{latexsym,bm}
\usepackage{geometry}

\def \be {\begin{equation}}
\def \ee {\end{equation}}
\def \bea {\begin{eqnarray}}
\def \eea {\end{eqnarray}}
\def \nn {\nonumber}

\def \a {\alpha}
\def \b {\beta}

\def \d {\delta}

\def \m {\mu}
\def \n {\nu}
\def \k {\kappa}

\def \s {\sigma}
\def \r {\rho}
\def \o {\omega}

\def \th {\theta}
\def \Th {\Theta}

\def \t {\tau}
\def \dag {\dagger}
\def \p {\partial}

\def\bd{\begin{document}}
\def\ed{\end{document}}
\def\nn{\nonumber}
\def\bea{\begin{eqnarray}}
\def\eea{\end{eqnarray}}
\let\bm=\bibitem
\let\la=\label

\def\N{{\cal N}}
\def\sst{\scriptscriptstyle}
\def\thetabar{\bar\theta}
\def\Tr{{\rm Tr}}
\def\one{\mbox{1 \kern-.59em {\rm l}}}

\def\a{\alpha}      \def\da{{\dot\alpha}}
\def\b{\beta}       \def\db{{\dot\beta}}
\def\c{\gamma}  \def\C{\Gamma}  \def\cdt{\dot\gamma}
\def\d{\delta}  \def\D{\Delta}  \def\ddt{\dot\delta}
\def\e{\epsilon}        \def\vare{\varepsilon}
\def\f{\phi}    \def\F{\Phi}    \def\vvf{\f}
\def\h{\eta}
\def\k{\kappa}
\def\l{\lambda} \def\L{\Lambda}
\def\m{\mu} \def\n{\nu}
\def\o{\omega}
\def\P{\Pi}
\def\r{\rho}
\def\s{\sigma}  \def\S{\Sigma}
\def\t{\tau}
\def\th{\theta} \def\Th{\Theta} \def\vth{\vartheta}
\def\X{\Xeta}
\def\z{\zeta}
\def\w{\wedge}
\def\u{\underline}
\def\hs{\hspace}

\def\cA{{\cal A}} \def\cB{{\cal B}} \def\cC{{\cal C}}
\def\cD{{\cal D}} \def\cE{{\cal E}} \def\cF{{\cal F}}
\def\cG{{\cal G}} \def\cH{{\cal H}} \def\cI{{\cal I}}
\def\cJ{{\cal J}} \def\cK{{\cal K}} \def\cL{{\cal L}}
\def\cM{{\cal M}} \def\cN{{\cal N}} \def\cO{{\cal O}}
\def\cP{{\cal P}} \def\cQ{{\cal Q}} \def\cR{{\cal R}}
\def\cS{{\cal S}} \def\cT{{\cal T}} \def\cU{{\cal U}}
\def\cV{{\cal V}} \def\cW{{\cal W}} \def\cX{{\cal X}}
\def\cY{{\cal Y}} \def\cZ{{\cal Z}}

\def\ua{\underline{\alpha}} \def\ubb{\underline{\beta}}
\def\ug{\underline{\gamma}}
\def\ub{\underline{\phantom{\alpha}}\!\!\!\beta}
\def\uc{\underline{\phantom{\alpha}}\!\!\!\gamma}
\def\um{\underline{\mu}} \def\un{\underline{\nu}}
\def\ud{\underline\delta}
\def\ue{\underline\epsilon}
\def\una{\underline a}\def\unA{\underline A}
\def\unb{\underline b}\def\unB{\underline B}
\def\unc{\underline c}\def\unC{\underline C}
\def\und{\underline d}\def\unD{\underline D}
\def\une{\underline e}\def\unE{\underline E}
\def\unf{\underline{\phantom{e}}\!\!\!\! f}\def\unF{\underline F}
\def\unm{\underline m}\def\unM{\underline M}
\def\unn{\underline n}\def\unN{\underline N}
\def\unp{\underline{\phantom{a}}\!\!\! p}\def\unP{\underline P}
\def\unq{\underline{\phantom{a}}\!\!\! q}
\def\unQ{\underline{\phantom{A}}\!\!\!\! Q}
\def\unH{\underline{H}}
\def\ul{\underline}

\def\As {{A \hspace{-6.4pt} \slash}\;}
\def\bs {{b \hspace{-6.4pt} \slash}\;}
\def\Ds {{D \hspace{-6.4pt} \slash}\;}
\def\ds {{\del \hspace{-6.4pt} \slash}\;}
\def\ss {{\s \hspace{-6.4pt} \slash}\;}
\def\ks {{ k \hspace{-6.4pt} \slash}\;}
\def\ps {{p \hspace{-6.4pt} \slash}\;}
\def\pas {{{p_1} \hspace{-6.4pt} \slash}\;}
\def\pbs {{{p_2} \hspace{-6.4pt} \slash}\;}

\def\Fh{\hat{F}}
\def\Vh{\hat{V}}
\def\Xh{\hat{X}}
\def\ah{\hat{a}}
\def\xh{\hat{x}}
\def\yh{\hat{y}}
\def\ph{\hat{p}}
\def\xih{\hat{\xi}}

\def\psit{\tilde{\psi}}
\def\Psit{\tilde{\Psi}}
\def\tht{\tilde{\th}}

\def\At{\tilde{A}}
\def\Qt{\tilde{Q}}
\def\Rt{\tilde{R}}
\def\Nt{\tilde{N}}

\def\at{\tilde{a}}
\def\st{\tilde{s}}
\def\ft{\tilde{f}}
\def\pt{\tilde{p}}
\def\qt{\tilde{q}}
\def\vt{\tilde{v}}
\def\nt{\tilde{n}}

\def\delb{\bar{\partial}}
\def\bz{\bar{z}}
\def\bD{\bar{D}}
\def\bB{\bar{B}}

\def\bk{{\bf k}}
\def\bl{{\bf l}}
\def\bp{{\bf p}}
\def\bq{{\bf q}}
\def\br{{\bf r}}
\def\bx{{\bf x}}
\def\by{{\bf y}}
\def\bR{{\bf R}}
\def\bV{{\bf V}}

\def\d{\delta}\def\D{\Delta}\def\ddt{\dot\delta}

\def\p{\partial} \def\del{\partial}
\def\xx{\times}
\def\uno{\mbox{1 \kern-.59em {\rm l}}}

\def\trp{^{\top}}
\def\inv{^{-1}}
\def\dag{{^{\dagger}}}

\def\pr{\prime}
\def\rar{\rightarrow}
\def\lar{\leftarrow}
\def\lrar{\leftrightarrow}

\title{$D_2$ Chern-Simons Gravity}
\author{
Bin Chen$^{1,2,3}$\footnote{bchen01@pku.edu.cn},\,
Jiang Long$^{1,2}$\footnote{lj301@pku.edu.cn}\,
and
Yi-Nan Wang$^{1}$\footnote{ynwang@pku.edu.cn}
}
\date{}

\begin{document}

\maketitle

\begin{center}
{\it
$^{1}$Department of Physics, Peking University, Beijing 100871, P.R. China\\
\vspace{2mm}
$^{2}$State Key Laboratory of Nuclear Physics and Technology, Peking University, Beijing 100871, P.R. China\\
\vspace{2mm}
$^{3}$Center for High Energy Physics, Peking University, Beijing 100871, P.R. China\\
}
\vspace{10mm}
\end{center}

\date{}


\maketitle

\begin{abstract}
We study the theory that contains two spin-2 fields. This theory can be regarded as a simplified version of higher spin gravity in AdS$_3$. It can be formulated either in the first order formulation or in the second order formulation. From the first order formulation we  construct the black holes in the theory and study the thermodynamics of these black holes with the help of dual holographic OPE. From the higher spin point of view, this black hole is most naturally regarded as a black hole with a spin-2 hair. In the second order formulation, we
obtain an action for the metric and extra spin-2 fields and analyze the corresponding black holes. Even though for some simple cases, the conventional notions, such as the horizon, may help us to read part of the information of the black hole, they break down in generic cases. On the other hand, the action in the second order formulation could be rewritten in the form of a decoupled AdS$_3$ bi-gravity. Moreover, the entropy of the black hole with the spin-2 hair could be reproduced from the AdS$_3$ bi-gravity in the Euclidean path-integral formalism.
 \end{abstract}
 \newpage
\section{Introduction}

The most remarkable feature of gravity is that it couples to all forms of matter universally. For the scalar, spin-1 gauge field, and antisymmetric tensor field\footnote{Fermions can also be introduced, however, we focus on the bosonic theory here.}, their minimal couplings with gravity are well-defined and the gauge transformations of these fields (if they have such) do not mix with the diffeomorphism. Therefore the conventional notions of spacetime remain the same as the ones in pure gravity. As a result, the action describing the gravity and its interaction with other fields is easily written in a second order formulation. In these cases, the first order formulation is possible, but the physical meaning is actually more straightforward in the second order formulation.


However, the situation differs completely in the so-called higher spin gravity theory. At the tree level, the free higher spin theory with a spin larger than two could be defined in any background\cite{Fronsdal1:1978, Fronsdal2:1978} in the second order formulation, as reviewed in \cite{Bouatta:2004}. Once the interaction is considered, the theory becomes complicated and messy in the second order formulation. For the most important case, the candidate theory is the so-called Vasiliev theory\cite{Vasiliev}, which is best developed in the first order formulation. There are a few remarkable properties on the Vasiliev's higher spin gravity. First of all, the theory is only well-defined in the spacetime background with a non-vanishing cosmological constant. Secondly, the theories in the spacetime larger than three involve an infinite number of higher spin, and moreover are only defined by the on-shell equations of motion without off-shell actions. In the simplest case, say, the minimal bosonic Vasiliev AdS$_4$ gravity, it still contains all even spins $s=0,2,4,\cdots,\infty$. Of course, in  three dimensions, the theory can have a finite truncation\cite{ASG II:2010} by using the Chern-Simons formulation along the lines of \cite{Blencowe:1989, Blencowe:1990}. In a recent investigation\cite{truncated}, a theory containing only  spin-2 and spin-6 has also been constructed, suggesting that the higher spin gravity could be defined with respect to every Lie algebra. The last but the most important property from the gravity point of view is that higher spin gauge transformation is always mixed with the diffeomorphism and hence changes the notions of  spacetime geometry in general relativity (GR) dramatically. So it is quite amazing that one can still analyze the asymptotic symmetry \cite{ASG I:2010,ASG II:2010},  construct the black holes with higher spin hair\cite{Per kraus:2011,Ammon:2011nk} and discuss their thermodynamics in the AdS$_3$ higher spin theory.

The original work on the spin-3 black hole\cite{Per kraus:2011} was actually guided by the dual CFT description and relied heavily on the Chern-Simons(first order) formulation. It showed explicitly that the conventional notions of GR, like the horizons and the singularity, are gauge dependent and make little sense. Nevertheless, the thermodynamics of a higher spin black hole is still well-defined by working with gauge invariant quantities. As in the first order formulation the physical Fronsdal fields do not appear explicitly, it is better to have a second order formulation to understand the whole physics. There have been several attempts on this issue\cite{Campoleoni:2012hp,{Ippei Fujisawa}}, but it is still far from accomplished. In particular, in \cite{Campoleoni:2012hp} the metric-like formulation of a spin-3 field minimally coupled to gravity has been
constructed perturbatively and the higher spin corrections to the black hole entropy have been computed using Wald's formula. 

Higher spin gauge theory is often regarded as a theory  halfway between gravity and string theory, and it has received much attention in the context of AdS/CFT. One of the most important proposals is the HS/O(N) duality conjectured in \cite{Klevanov} and nicely reviewed in \cite{Giombi:2012he}. Hence one can pose the question  whether there is a theory which is  halfway between the Einstein gravity and the higher spin gravity. From the Einstein gravity point of view, the next theory we need to consider is actually the one of a spin-2 field coupling to gravity by simply increasing the spin from $0,1$ to $2$. From the higher spin gauge theory, the theory we need to consider is also  the one of a spin-2 field coupling to gravity by simply decreasing the higher spin to $2$. Here is a ``critical'' point where the usual Einstein gravity meets the higher spin gravity. We expect that such kind of theory inherits many properties from both sides and hence provides a better understanding of both the Einstein gravity and the higher spin gravity. This turns out to be the case. The theory we construct is a spin-$2$ field coupled to the AdS$_3$ gravity. It can be regarded as an extension of the Maxwell theory coupling to gravity. However, we will find that the spin-$2$ matter field has its own gauge transformation and such kind of gauge transformation mixes with the diffeomorphism. This is a feature that all higher spin gauge theories share. Namely, it changes the notions of spacetime geometry, but not so dramatically as in the higher spin gauge theory, hence it is still under control in the sense that the number of the matter field is finite. More interestingly, the theory can be constructed both in the first order formalism and the second order formalism.  Furthermore, we can identify the black holes in the theory and study their thermodynamics in two different formulations. This may shed light on the study of thermodynamics of the higher spin black holes. 

Though our work is motivated from the study of the higher spin gravity and holography, the final theory we construct  is closely related to the so-called bi-gravity or multi-gravity\footnote{For a recent generalization, see\cite{Hassan:2011, Hinterbichler:2012}.}\cite{salam}. However, in the field of bi-gravity, the attention has been focused on searching for ghost free theories after introducing interaction between different gravitons. According to a no-go theorem\footnote{This no-go theorem could be evaded in the  context of higher spin gravity, as the Vasiliev theory itself allows for a supersymmetric or Chan-Paton factor\cite{Prokushkin:1998bq} generalization which includes multiple massless gravitons. These generalizations are important for the recent proposed duality given in \cite{Chang:2012kt}.}
in \cite{Henneaux:2000}, the interaction between different gravitons must induce mass terms to some gravitons, hence in a theory with multiple massless gravitons the gravitons are decoupled with each other. This kind of theory has been ignored in the field of bi-gravity. On some aspects, we revisit this problem and find that this decoupled theory could be related to  the simplest example of higher spin gravity with a spin-2 matter field and a massless graviton. We can identify the black holes in the bi-gravity theory and study their thermodynamics from their holographic duals. As far as we know, these  kinds of non-perturbative objects have never been fully understood in bi-gravity. One lesson from our study is that in such decoupled bi-gravity theory, the two massless gravitons seem to live in their own world, and have their own non-perturbative excitations. As a result the entropy of the black hole with spin-2 hair in the original theory could be reproduced successfully
from the sum of the entropies of BTZ black holes in bi-gravity, with each BTZ black hole as the solution of independent AdS$_3$ gravity.


The structure of the paper is as follows. In section 2, we introduce our theory in the first order formalism, including its action, symmetry and spectrum. We construct the black holes in this theory and fully explore their thermodynamic properties. In section 3, we go to the second order formalism, and discuss the properties using the conventional methods. In particular we show how to reproduce the entropy of the black hole from bi-gravity point of view.  In section 4, we give the dual picture of our theory. We close the paper with some discussions and the conclusion.

\section{First Order Formulation}

In this section, we discuss our model in the first order formulation and show its relation with the higher spin gravity with the gauge group $SO(2,2)$. We construct the black hole solution with an extra spin-2 hair and study its thermodynamics.

\subsection{Action and spectrum}

Let us consider the following action
\be
I=\frac{k}{4\pi}\int tr'(AdA+\frac{2}{3}A^3+2AA'A'+A'dA')-bar\  term.\label{action}
\ee
This action is invariant under the gauge transformations
\bea
\delta A=d\epsilon+[A,\epsilon]+[A',\epsilon'],\nn\\
\delta A'=d\epsilon'+[A,\epsilon']+[A',\epsilon]\label{symm}
\eea
up to boundary terms. Here $k$ is the Chern-Simons level and $A$,$A'$ takes value in suitable Lie algebra. Note that when $A'=0$, the system reduces to the familiar higher spin gravity in AdS$_3$. To relate the theory to gravity, we need to identify the level $k$ and the Newton constant $G$ by the following relation\footnote{Our convention of the trace operation is as follows, a simple $tr$ means the trace of the matrix, the $tr'$ in the action is chosen to preserve the relation between k and G.}
\be
k=\frac{l}{4G}
\ee
where $l$ is the AdS radius. We also introduce the vielbein and the spin connection through
\bea
A&=&\omega+\frac{e}{l},\nn\\
\bar{A}&=&\omega-\frac{e}{l},\nn\\
A'&=&\omega'+\frac{e'}{l},\nn\\
\bar{A}'&=&\omega'-\frac{e'}{l}.
\eea
To analyze the spectrum in this theory, we  consider the equations of motion and expand them around the vacuum. The equations of motion turn out to be
\bea
dA+A\wedge A+A'\wedge A'=0,\nn\\
dA'+A\wedge A'+A'\wedge A=0.\label{*2}
\eea
We choose the gauge algebra to be $sl(2,R)$ for $A,\bar{A},A',\bar{A}'$. Obviously, we can take the vacuum to be AdS$_3$\footnote{From now on, we set the AdS radius $l$ to be 1.}
\bea
A_{AdS}'&=&0,\nn\\
\bar{A}_{AdS}'&=&0,\nn\\
A_{AdS}&=&(e^{\rho}J_1+\frac{1}{4}e^{-\rho}J_{-1})dx^++J_0d\rho,\\
\bar{A}_{AdS}&=&-(e^{\rho}J_{-1}+\frac{1}{4}e^{-\rho}J_1)dx^--J_0d\rho,\nn
\eea
where we have introduced the $sl(2,R)$ generators $J_{-1},J_0,J_1$ to be
\begin{align}
&J_1=\left(\begin{array}{cc}
0&0\\-1&0\end{array}\right),\hs{3ex}J_0=\frac{1}{2}\left(\begin{array}{cc}
1&0\\0&-1\end{array}\right),\hs{3ex} J_{-1}=\left(\begin{array}{cc}
0&1\\0&0\end{array}\right).\end{align}
Expanding the equations of motion to linearized order,  we find
\bea
da+A\w a+a\w A=0,\nn\\
da'+A\w a'+a'\w A=0,\nn\\
d\bar{a}+\bar{A}\w \bar{a}+\bar{a}\w \bar{A}=0,\label{eom}\\
d\bar{a}'+\bar{A}\w \bar{a}'+\bar{a}'\w \bar{A}=0\nn
\eea
where we still denote the background as $A,\bar{A}$ for simplicity. The $a,\bar{a},a',\bar{a}'$ denote the fluctuations of $A,\bar{A},A',\bar{A}'$ respectively.
Note that the equations (\ref{eom}) have been analyzed in \cite{Hstmg}, where it has been shown that the spectrum only depends on the commutation relations of the generators. In the case at hand, the spectrum just consists of two massless spin-2 fluctuations. For the general case, if $A,\bar{A},A',\bar{A}'$ takes value in $sl(n,R)$, the spectrum would consist of two copies of massless spin $s=2,3,\cdots,n$. Moreover, as the recent work \cite{truncated} suggested, we can take other gauge groups such as $Sp(2n,R),SO(2n+1,R)$ and $G_2$, the spectrum would change accordingly.

\subsection{Black holes}

It is not {\it a priori} evident that the theory we defined above can have consistent black hole solutions. In this subsection, we show that we can indeed construct the black holes along the work of \cite{Per kraus:2011,Ammon:2011nk} on the higher spin black holes. Still, for simplicity, we choose the gauge group to be $SL(2,R)$.

The first step is to rewrite the action to be
\be
I=\frac{k}{8\pi}\int tr' [(A+A')d(A+A')+\frac{2}{3}(A+A')^3+(A-A')d(A-A')+\frac{2}{3}(A-A')^3]-bar\ term.\label{action2}
\ee
This is nothing but two copies of conventional higher spin gravity. More compactly, with a $B$(and $\bar{B}$ correspondingly) of the form
\be
B=\left(
\begin{array}{cc}
A+A'&0\\0&A-A'
\end{array}
\right),
\ee
the action can be rewritten as\footnote{It seems that there is a factor $\frac{1}{2}$ in the action, however, we can always absorb it into the definition of the trace.}
\be
I=S_{CS}[B]-S_{CS}[\bar{B}].\label{B action}
\ee
The equations of motion become
\be
dB+B\w B=0,\ d\bar{B}+\bar{B}\w \bar{B}=0.
\ee
The gauge transformations are now
\be
\delta B=d\lambda+[B,\lambda],\hs{3ex}\delta\bar{B}=d\bar{\lambda}+[\bar{B},\bar{\lambda}]
\ee
where the gauge parameters $\lambda,\bar{\lambda}$ are related to the original gauge parameters $\epsilon,\bar{\epsilon}$ by
\be
\lambda=\left(
\begin{array}{cc}
\epsilon+\epsilon'&0\\0&\epsilon-\epsilon'
\end{array}
\right),\ \bar{\lambda}=\left(
\begin{array}{cc}
\bar{\epsilon}+\bar{\epsilon}'&0\\0&\bar{\epsilon}-\bar{\epsilon}'
\end{array}
\right).
\ee
The theory is just a higher spin gravity based on the gauge group $SL(2,R)\times SL(2,R) \simeq SO(2,2)$.  Based on the arguments in \cite{truncated}, this gauge algebra is $D_2$ and hence has rank 2, and the corresponding higher spin gravity describes another spin-$2$ field interacting with the gravity.

We can construct the black hole with an extra spin-$2$ hair in this theory and analyzed its thermodynamics explicitly with the help of dual CFT.  At first, we need to introduce the generators of $D_2$. Let us  introduce six generators $L_{-1},L_0,L_1,V_{-1},V_0,V_1$ as
\begin{align}
&L_i=\left(\begin{array}{cc}
J_i&0\\0&J_i
\end{array}\right), \hs{3ex}V_i=\left(\begin{array}{cc}
J_i&0\\0&-J_i\end{array}\right),
\end{align}
which have the following commutation relations
\be
[L_i,L_j]=(i-j)L_{i+j},\hs{3ex}[L_i,V_j]=(i-j)V_{i+j},\hs{3ex}[V_i,V_j]=(i-j)L_{i+j},
\ee
with $i=-1,0,1$. Obviously the generators $L_i$ form an $sl(2,R)$ subalgebra and correspond to the
graviton, while the $V_i$'s correspond to the extra spin-$2$ field.

The black hole solution can be constructed directly by solving the flatness equations. To have a spin-2 hair,
we make the following ansatz:
\bea
B&=&(e^{\rho}L_1-\frac{2\pi}{k}\mathcal{L}L_{-1}e^{-\rho}-\frac{2\pi}{k}\mathcal{L}'e^{-\rho}V_{-1})dx^+\nn\\
 && +q(-\frac{2\pi}{k}\mathcal{L}' L_{-1}e^{-\rho}+e^{\rho}V_1-\frac{2\pi}{k}\mathcal{L}e^{-\rho}V_{-1})dx^-+L_0d\rho,  \label{BH}\\
\bar{B}&=&-(e^{\rho}L_{-1}-\frac{2\pi}{k}\bar{\mathcal{L}}L_{1}e^{-\rho}-\frac{2\pi}{k}\bar{\mathcal{L}}'e^{-\rho}V_{1})dx^-\nn\\
& &-\bar{q}(-\frac{2\pi}{k}\bar{\mathcal{L}}' L_{1}e^{-\rho}+e^{\rho}V_{-1}-\frac{2\pi}{k}\bar{\mathcal{L}}e^{-\rho}V_{1})dx^+-L_0d\rho.\nn
\eea
As we will show, this solution can be interpreted as a black hole. In it, the parameter $\mathcal{L}$ should be interpreted as the charge of the graviton, encoding the information of the mass and the angular momentum,  and ${\mathcal{L}}'$ should be interpreted as the other spin-2 charge. The parameter $q$ should be interpreted as the chemical potential corresponding to the spin-2 charge ${\mathcal{L}}'$. The interpretation of the parameters with a bar is similar. To see this, we first turn off the chemical potentials $q$ and $\bar{q}$, and read the physical meaning of $\mathcal{L}$ and ${\mathcal{L}}'$ from the asymptotic symmetry analysis\cite{ASG II:2010}. We can make a gauge transformation to set the field $B$ to be
\be
B=e^{-L_0\rho}(b+d)e^{L_0\rho}
\ee
where
\be
b=(L_1-\frac{2\pi}{k}\mathcal{L}(x^+)L_{-1}-\frac{2\pi}{k}\mathcal{L}'(x^+)V_{-1})dx^+,
\ee and we allow $\mathcal{L}$ and $\mathcal{L}'$ to be the functions of $x^+$. An arbitrary gauge parameter $\lambda$ can be parameterized as
\be
\lambda=\lambda_1 L_1+\lambda_0L_0+\lambda_{-1}L_{-1}+\xi_1 V_1+\xi_0 V_0+\xi_{-1}V_{-1}.
\ee
The gauge transformation that preserves the asymptotically $AdS_3$ boundary condition can be solved to be
\bea
\lambda_0=-\partial\lambda,\hs{3ex} \lambda_{-1}=\frac{1}{2}\partial^2\lambda-\frac{2\pi}{k}\mathcal{L}\lambda-\frac{2\pi}{k}\mathcal{L}'\xi,\nn\\
\xi_0=-\partial\xi,\hs{3ex} \xi_{-1}=\frac{1}{2}\partial^2\xi-\frac{2\pi}{k}\mathcal{L}\xi-\frac{2\pi}{k}\mathcal{L}'\lambda,
\eea
where $\partial$ means $\partial_+$ and we still denote $\lambda=\lambda_1,\xi=\xi_1$ without causing confusion. Under the gauge transformation, $\mathcal{L}$ and $\mathcal{L}'$ transform   as
\bea
-\delta\mathcal{L}=\frac{k}{4\pi}\partial^3\lambda-\partial\mathcal{L}\lambda-2\mathcal{L}\partial{\lambda}-2\mathcal{L}'\partial\xi-\partial\mathcal{L}'\xi,\nn\\
-\delta\mathcal{L}'=-\partial\mathcal{L}'\lambda-2\mathcal{L}'\partial\lambda-\partial\mathcal{L}\xi-2\mathcal{L}\partial\xi+\frac{k}{4\pi}\partial^3\xi.
\eea
If we identify the stress tensor $\mathcal{T}=-2\pi\mathcal{L}$ and the spin-2 charge $\tilde{\mathcal{T}}=-2\pi\mathcal{L}'$, then the above transformations can be changed to the following OPE
\bea
\mathcal{T}(z)\mathcal{T}(0)&\sim&\frac{c/2}{z^4}+\frac{2\mathcal{T}}{z^2}+\frac{\partial\mathcal{T}}{z},\nn\\
\mathcal{T}(z)\tilde{\mathcal{T}}(0)&\sim&\frac{2\tilde{\mathcal{T}}}{z^2}+\frac{\partial\tilde{\mathcal{T}}}{z},\nn\\
\tilde{\mathcal{T}}(z)\tilde{\mathcal{T}}(0)&\sim&\frac{c/2}{z^4}+\frac{2\mathcal{T}}{z^2}+\frac{\partial\mathcal{T}}{z}.\label{OPE}
\eea
The first OPE is just the conventional Virasoro algebra with central charge $c=6k$, the second OPE tells us that the $\tilde{\mathcal{T}}$ is a primary operator with the conformal weight (2,0). This is the reason why we should identify $\mathcal{T}$ as the boundary stress tensor and $\tilde{\mathcal{T}}$ as the spin-2 current. Actually, an arbitrary (nonzero) rescaling of $\tilde{\mathcal{T}}$ still gives us the primary operator with the conformal weight (2,0).

To understand  the parameters $q$ and $\bar{q}$ as the chemical potentials corresponding to $\mathcal{L}'$ and $\bar{\mathcal{L}}'$ , we need to analyze the Ward identity. By adding an operator $exp^{\int \mu' \mathcal{L}'}$ to the boundary CFT, we should change the ansatz on $b$
\be
b=(L_1-\frac{2\pi}{k}\mathcal{L}L_{-1}-\frac{2\pi}{k}\mathcal{L}'V_{-1})dx^++(qV_1+q_0V_0+q_{-1}V_{-1}+\nu L_{-1})dx^-,
\ee
where we allow $\mathcal{L},\mathcal{L}',q,q_0,q_{-1},\nu$ to be the functions of $x^+$ and $x^-$. Then by solving the flatness equations we find that
\bea
\partial_-\mathcal{L}&=&q\partial_+\mathcal{L}'+2\partial_+q\mathcal{L}',\nn\\
\partial_-\mathcal{L}'&=&2\partial_+q\mathcal{L}+q\partial_+\mathcal{L}-\frac{k}{4\pi}\partial_+^3q.\label{WI}
\eea
We take the values of $\mathcal{L},\mathcal{L}'$ as the expectation values when  an operator $exp^{\int \mu' \mathcal{L}'}$ is inserted into the boundary CFT, namely
\be
\mathcal{L}=<\mathcal{L}>_{\mu'},\ \mathcal{L}'=<\mathcal{L}'>_{\mu'}
\ee
where the subscript $\mu'$ means that for arbitrary operator $\mathcal{O}$,
\be
<\mathcal{O}>_{\mu'}=<\mathcal{O} exp^{\int \mu'\mathcal{L}'}>.
\ee
By choosing $\mathcal{O}$ to be $\mathcal{L},\mathcal{L}'$ and calculating\footnote{Here $\partial_{-}$ means $\partial_{\bar{z}}$, and the rule from Lorentz signature to Euclidean signature is $x^+\to z, x^-\to-\bar{z}$.} $\partial_-<\mathcal{L}>_{\mu'},\partial_-<\mathcal{L}'>_{\mu'}$ and comparing the result with (\ref{WI}), we find that we should identify
\be
q=-\mu',\hs{3ex}\mathcal{L}=-\frac{\mathcal{T}}{2\pi}.
\ee
The first identity shows us the physical meaning of $q$: it is the chemical potential conjugate to the spin-2 charge $\mathcal{L}'$. The second identity is consistent with the OPE result found before. This completes our discussion on the parameters appearing in the solution.

The next problem is to study the thermodynamics of the black hole, which has a metric and a spin-2 hair.
From now on we rescale $\frac{2\pi}{k}\mathcal{L}\to\mathcal{L},\frac{2\pi}{k}\mathcal{L}'\to\mathcal{L}'$ to simplify notation. The bar terms are also rescaled by the same factor. We can define the quantities $\tau,\bar{\tau},\alpha,\bar{\alpha}$ by
\be
\tau=\frac{i}{2\pi k}\frac{\partial S}{\partial\mathcal{L}},\ \alpha=\frac{i}{2\pi k}\frac{\partial S}{\partial\mathcal{L}'},\ \bar{\tau}=-\frac{i}{2\pi k}\frac{\partial S}{\partial\bar{\mathcal{L}}},\ \bar{\alpha}=-\frac{i}{2\pi k}\frac{\partial S}{\partial\bar{\mathcal{L}}'},
\ee
where $S$ is the entropy of the black hole and is a function of $\mathcal{L},\mathcal{L}',\bar{\mathcal{L}},\bar{\mathcal{L}'}$. The
$\alpha,\bar{\alpha}$ are the potentials conjugate to $\mathcal{L}',\bar{\mathcal{L}}'$ which appear in the partition function
\be
Z=Tr \exp(i2\pi k(\tau\mathcal{L}+\alpha\mathcal{L}'-\bar{\tau}\bar{\mathcal{L}}-\bar{\alpha}\bar{\mathcal{L}}')),
\ee
and are related to the chemical potentials $\mu'$ and $\bar{\mu}'$  by $\bar{\tau}\mu'=\alpha,\tau\bar{\mu}'=\bar{\alpha}$. The holonomy is $\omega=2\pi(\tau b_+-\bar{\tau}b_-)$.
As in the study of other higher spin black holes, we require that the black hole have the same holonomy as the BTZ black hole. For the BTZ black hole, the holonomy now is $\pm i\pi,\pm i\pi$. Since the gauge group is rank 2, we need to solve two independent holonomy equations. We choose the following two equations
\be
P_2=tr\omega^2=-4\pi^2,\ P_4=tr\omega^4=4\pi^4.
\ee
We require the solution to have a BTZ limit when the spin-2 charge vanishes. Considering the dimensions of $\mathcal{L},\mathcal{L}'$, we assume the general form of the entropy  to be
\be
S=2\pi k\sqrt{\mathcal{L}}f(y)
\ee
with $y=\frac{\mathcal{L}'}{\mathcal{L}}$ being dimensionless. Taking all these into account, we  find that the holonomy constraints become the following two differential equations
\bea
1&=&f(y)^2 - 4 (-1 + y^2)f(y)'^2\nn\\
2&=&(1 + y)^2 (f(y) - 2 (-1 + y)f(y)')^4 + (-1 + y)^2 (f(y) - 2 (1 + y)f(y)')^4
\eea
These two equations plus the boundary condition $f(0)=1$ can be satisfied by the solution
\be
f(y)=\cos\frac{1}{2}\arcsin y
\ee
where $-1\leq y\leq 1$. Hence
the black hole has a consistent first law of thermodynamics with the entropy as
\be
S=2\pi k\sqrt{\mathcal{L}}\cos(\frac{1}{2}\arcsin{y}).\label{entropy}
\ee
Taking the bar term into account, we find that the entropy could be written into a more suggestive form
\be
S=\pi k(\sqrt{\mathcal{L}+\mathcal{L}'}+\sqrt{\mathcal{L}-\mathcal{L}'}+\sqrt{\bar{\mathcal{L}}+\bar{\mathcal{L}}'}+\sqrt{\bar{\mathcal{L}}-\bar{\mathcal{L}}'}).\label{entropy'}
\ee

The extreme cases correspond to $y=\pm1$, namely $\mathcal{L}'=\pm \mathcal{L}$ and $\bar{\mathcal{L}}'=\pm \bar{\mathcal{L}}$. It is interesting to see that the spin-2 charge could be negative, but its sign does not
affect the entropy as the function $f(y)$ is an even function of $y$. This fact can also be seen from the entropy formula (\ref{entropy'}). This is very different from the black hole with spin-4 or spin-6 hair, where the entropy functions are sensitive to the sign of the higher spin charges\cite{truncated}.  If the spin-2 charge is zero, we simply have a BTZ black hole, which could be described by a boundary CFT with central charge $6k$ and at the levels $k\mathcal{L}, k\bar{\mathcal{L}}$. If the spin-2 charge is nonvanishing, the black hole entropy gets contributions from the spin-2 fields.

Some relevant quantities can be computed as
\bea
\tau=\frac{i}{4}(\frac{1}{\sqrt{\mathcal{L}+\mathcal{L}'}}+\frac{1}{\sqrt{\mathcal{L}-\mathcal{L}'}}),\ \alpha=\frac{i}{4}(\frac{1}{\sqrt{\mathcal{L}+\mathcal{L}'}}-\frac{1}{\sqrt{\mathcal{L}-\mathcal{L}'}}),\\
\bar{\tau}=-\frac{i}{4}(\frac{1}{\sqrt{\bar{\mathcal{L}}+\bar{\mathcal{L}}'}}+\frac{1}{\sqrt{\bar{\mathcal{L}}-\bar{\mathcal{L}}'}}),\ \bar{\alpha}=-\frac{i}{4}(\frac{1}{\sqrt{\bar{\mathcal{L}}+\bar{\mathcal{L}}'}}-\frac{1}{\sqrt{\bar{\mathcal{L}}-\bar{\mathcal{L}}'}}).
\eea
The partition function turns out to be
\be
\ln Z=\frac{\pi k}{2}(\sqrt{\mathcal{L}+\mathcal{L}'}+\sqrt{\mathcal{L}-\mathcal{L}'}+\sqrt{\bar{\mathcal{L}}+\bar{\mathcal{L}}'}+\sqrt{\bar{\mathcal{L}}-\bar{\mathcal{L}}'}).\label{partion1}
\ee
For the following discussion, we also give the value of $q$
\be
q=\sqrt{\frac{(\bar{\mathcal{L}}+\bar{\mathcal{L}}')(\bar{\mathcal{L}}-\bar{\mathcal{L}}')}{(\mathcal{L}+\mathcal{L}')(\mathcal{L}-\mathcal{L}')}}(\frac{\sqrt{\mathcal{L}-\mathcal{L}'}-\sqrt{\mathcal{L}+\mathcal{L}'}}{\sqrt{\bar{\mathcal{L}}-\bar{\mathcal{L}}'}+\sqrt{\bar{\mathcal{L}}+\bar{\mathcal{L}}'}}).\label{q}
\ee

\section{Second Order Formulation}

In this section, we give an alternative derivation of the previous results in the second order formulation.

\subsection{Action and spectrum}
 The action (\ref{action}) can be written as (\ref{action2})
\be
S=\frac{k}{8\pi}\int tr' [(A+A')d(A+A')+\frac{2}{3}(A+A')^3+(A-A')d(A-A')+\frac{2}{3}(A-A')^3]-bar\ term \nn
\ee
Hence, we can define two tensor fields as
\be
g_{\mu\nu}^{(1)}=2 tr(e_{\mu}+e'_{\mu})(e_{\nu}+e'_{\nu}),\ g_{\mu\nu}^{(2)}=2 tr(e_{\mu}-e'_{\mu})(e_{\nu}-e'_{\nu}).
\ee
Since the Chern-Simons action
\be
I=\frac{k}{4\pi}\int tr' (CdC+\frac{2}{3}C^3)-bar\ term
\ee
can be written in the second order formulation as
\be
I=\frac{1}{16\pi G}\int d^3x \sqrt{-g}(R+2)
\ee
with the metric $g_{\mu\nu}=2tr(C_{\mu}C_{\nu})$, we find that the action (\ref{action2}) can be rewritten in the second order formulation as
\be
I=\frac{1}{32\pi G}\int d^3x \sqrt{-g^{(1)}}(R^{(1)}+2)+\frac{1}{32\pi G}\int d^3x \sqrt{-g^{(2)}}(R^{(2)}+2)\label{action3}
\ee
where $g^{(i)}$ is the determinant of the field $g^{(i)}_{\mu\nu}$ and $R^{(i)}$ is the Ricci scalar defined with respect to  the field $g^{(i)}_{\mu\nu}$. Note that the coefficient in front of the integral is half of the conventional Einstein gravity. This is inherited from the action (\ref{action2}).

It is remarkable that the action (\ref{action3}) is in accord with the no-go theorem in \cite{Henneaux:2000}, which states that multiple massless gravitons should decouple with each other, if we take $g^{(1)}_{\mu\nu}$
and $g^{(2)}_{\mu\nu}$ as the gravitons. In the original proof in \cite{Henneaux:2000} the discussion was focused on the fluctuations around a flat spacetime, while now the spacetime carries a negative cosmological constant. 


In our case, we have a well-defined theory, suggested from the first order formulation of the higher spin gravity theory. We have an interpretation to the action (\ref{action3}).
From the discussion in the last section, we identify $A$ to generate the spacetime $g_{\mu\nu}$ and $A'$ to generate the other spin-2 field $h_{\mu\nu}$. More explicitly, we have
\be
g_{\mu\nu}=2tr(e_{\mu}e_{\nu}+e'_{\mu}e'_{\nu}),\ h_{\mu\nu}=2tr(e_{\mu}e'_{\nu}+e_{\nu}e'_{\mu}),
\ee
which are related to $g^{(i)}_{\mu\nu}$ by
 \be
g^{(1)}_{\mu\nu}=g_{\mu\nu}+h_{\mu\nu},\ g^{(2)}_{\mu\nu}=g_{\mu\nu}-h_{\mu\nu}.\label{redef}
\ee
For us the action (\ref{action3}) actually describes an AdS$_3$ graviton field $g_{\mu\nu}$ coupling with a spin-2 matter field $h_{\mu\nu}$. Note that even though the action (\ref{action3}) is of an elegant form in terms of $g^{(i)}_{\mu\nu}$, it is really messy when written in terms of $g_{\mu\nu}$ and $h_{\mu\nu}$.

To give a consistency check of the action (\ref{action3}), we will show the following three points. Firstly, the action (\ref{action3}) has the same symmetry as before. Secondly, the action leads to the same spectrum as before, namely, a massless spin-2 graviton, and a massless spin-2 as matter. Lastly, we will show that the equations of motion induced from action (\ref{action3}) can be satisfied by the previous black hole solution.

Let us first consider the symmetry of the theory. A naive consideration suggests that the action (\ref{action3}) is only invariant under general coordinate transformations and hence there are some mismatches between this symmetry and the symmetry in the first order analysis (\ref{symm}), which indicates that each $g^{(i)}_{\mu\nu}$ should have its own diffeomorphism. However, a detailed consideration will show that the action (\ref{action3}) has a larger symmetry, in match with the first order result. The point is that two parts in (\ref{action3}) indeed transform separately. To see this, let us consider the transformation
\be
\delta_{\xi} g^{(1)}_{\mu\nu}=\nabla^{(1)}_{(\mu}\xi_{\nu)}.\label{trans1}
\ee
Since $R^{(1)}$ is constructed from $g^{(1)}$, we have
\be
R^{(1)\rho}_{\ \ \ \  \sigma\mu\nu}=\partial_{\mu}\Gamma^{(1)\rho}_{\ \ \ \sigma\nu}-\partial_{\nu}\Gamma^{(1)\rho}_{\ \ \ \sigma\mu}+\Gamma^{(1)\rho}_{\ \ \ \mu\lambda}\Gamma^{(1)\lambda}_{\ \ \ \sigma\nu}-\Gamma^{(1)\rho}_{\ \ \ \nu\lambda}\Gamma^{(1)\lambda}_{\ \ \ \sigma\mu}
\ee
\be
R^{(1)}_{\sigma\nu}=R^{(1)\mu}_{\ \ \ \ \sigma\mu\nu}, \hs{3ex}R^{(1)}=g^{(1)\sigma\nu}R^{(1)}_{\sigma\nu}
\ee
and hence
\be
\delta_{\xi} R^{(1)}=\xi^{\mu}\nabla^{(1)}_{\mu}R^{(1)},
\ee
then we find that
\be
\delta_{\xi}\int d^3x \sqrt{-g^{(1)}}(R^{(1)}+2)=\int d^3x \sqrt{-g^{(1)}} \nabla^{(1)}_{\mu}((R^{(1)}+2)\xi^{\mu}).
\ee
It is not evident that the right-hand side is just a boundary integral. But it can be shown that it is indeed a boundary term\footnote{See Appendix for an illustration.} and hence the action is invariant under the transformation (\ref{trans1}). Note that this kind of transformation is independent of the second part of the action and hence there is a second transformation on the second part of the action as well. In all, the action (\ref{action3}) indeed has a larger symmetry than coordinate transformation, in agreement with  what the first order formalism tells us.

The next step is to identify the spectrum of the system. Since the spin-2 field $h_{\mu\nu}$ is a matter field, it is suitable to let it vanish in the vacuum solution. Thus, the action (\ref{action3}) becomes the Einstein gravity with a negative cosmological constant in three dimensions. Then, the vacuum can be chosen to be AdS$_3$.
\be
g_{\mu\nu}=g^{AdS}_{\mu\nu},\hs{3ex} h_{\mu\nu}=0
\ee
 Let us expand the action around the vacuum to the second order to get the following linearized action
 \bea
 S&\sim&\frac{1}{32\pi G}\int \sqrt{-g}(-\frac{1}{2}\nabla_{\lambda}s^{(1)}\nabla_{\sigma}s^{(1)\lambda\sigma}+\frac{1}{2}\nabla_{\mu}s^{(1)\mu\nu}\nabla^{\lambda}s^{(1)}_{\lambda\nu}+\frac{1}{4}\nabla_{\lambda}s^{(1)}\nabla^{\lambda}s^{(1)}\nn\\
 &&-\frac{1}{4}\nabla_{\rho}s^{(1)}_{\mu\nu}\nabla^{\rho}s^{(1)\mu\nu}+\frac{1}{2}s^{(1)}_{\mu\nu}s^{(1)\mu\nu}+\frac{1}{2}(s^{(1)})^2+(1\to 2))+\mathcal{O}(s^3)\label{qua}
 \eea
 where $s^{(i)}$ denotes the fluctuation of $g^{(i)}$. Note that we can use the gauge symmetry discussed above to set the fluctuation to be transverse and traceless  and then get the linearized equations of motion of two massless spin-2 fields in the AdS$_3$ vacuum
 \be
 (\Box+2) s^{(i)}_{\mu\nu}=0.
 \ee
 This demonstrates that the spectrum of the theory in (\ref{action3}) consists of  two massless spin-2 fields. Note that the gravitational fluctuation is $s^{(1)}+s^{(2)}$ while the spin-2 matter fluctuation is $s^{(1)}-s^{(2)}$. It deserves a mention  that the quadratic action (\ref{qua}) allows a linearized gauge transformation $\delta s^{(i)}_{\mu\nu}=\nabla_{(\mu}\xi^{(i)}_{\nu)}$ which is in accordance with the linearized symmetry of the equations of motion (\ref{eom}).

 The last point which we want to check is that the solution (\ref{BH}) we obtained actually satisfies the equations of motion of the action (\ref{action3}). Let us first rewrite the solution (\ref{BH}) in terms of $A,A',\bar{A},\bar{A}'$ as\footnote{Be careful about the rescaling of $\mathcal{L},\mathcal{L}',\bar{\mathcal{L}},\bar{\mathcal{L}}'$.}
 \bea
A+A'&=&(e^{\rho}J_1-e^{-\rho}(\mathcal{L}+\mathcal{L}')J_{-1})dx^++
(qe^{\rho}J_1-q(\mathcal{L}+\mathcal{L}')e^{-\rho}J_{-1})dx^-+J_0d\rho\nn\\
A-A'&=&(e^{\rho}J_1-e^{-\rho}(\mathcal{L}-\mathcal{L}')J_{-1})dx^++(-qe^{\rho}J_1+qe^{-\rho}(\mathcal{L}-\mathcal{L}')J_{-1})dx^-+J_0d\rho\nn\\
\bar{A}+\bar{A}'&=&-(e^{\rho}J_{-1}-e^{-\rho}(\bar{\mathcal{L}}+\bar{\mathcal{L}'})J_1)dx^--(\bar{q}e^{\rho}J_{-1}-\bar{q}(\bar{\mathcal{L}}+\bar{\mathcal{L}'})e^{-\rho}J_1)dx^+-J_0d\rho\nn\\
\bar{A}-\bar{A}'&=&-(e^{\rho}J_{-1}-e^{-\rho}(\bar{\mathcal{L}}-\bar{\mathcal{L}'})J_1)dx^--(-\bar{q}e^{\rho}J_{-1}+\bar{q}e^{-\rho}(\bar{\mathcal{L}}-\bar{\mathcal{L}'})J_1)dx^+-J_0d\rho\nn
\eea
Then from the definition of $g^{(i)}$ we find
\bea
ds_1^2&=&(e^{\rho}-\bar{q}(\bar{\mathcal{L}}+\bar{\mathcal{L}'})e^{-\rho})(e^{-\rho}(\mathcal{L}+\mathcal{L}')-\bar{q}e^{\rho})(dx^+)^2\nn\\
&&-(qe^{\rho}-e^{-\rho}(\bar{\mathcal{L}}+\bar{\mathcal{L}'}))(e^{\rho}-q(\mathcal{L}+\mathcal{L}')e^{-\rho})(dx^-)^2\nn\\
&&+\left\{-(e^{\rho}-\bar{q}(\bar{\mathcal{L}}+\bar{\mathcal{L}'})e^{-\rho})(e^{\rho}-q(\mathcal{L}+\mathcal{L}')e^{-\rho})\right.\nn\\
&&\left.+(e^{-\rho}(\mathcal{L}+\mathcal{L}')-\bar{q}e^{\rho})(qe^{\rho}-e^{-\rho}(\bar{\mathcal{L}}+\bar{\mathcal{L}'}))\right\}dx^+dx^-+d\rho^2\label{me1}\\
ds^2_2&=&(e^{\rho}+\bar{q}(\bar{\mathcal{L}}-\bar{\mathcal{L}'})e^{-\rho})(e^{-\rho}(\mathcal{L}-\mathcal{L}')+\bar{q}e^{\rho})(dx^+)^2\nn\\
&&+(qe^{\rho}+e^{-\rho}(\bar{\mathcal{L}}-\bar{\mathcal{L}'}))(e^{\rho}+qe^{-\rho}(\mathcal{L}-\mathcal{L}'))(dx^-)^2\nn\\
&&+\left\{-(e^{\rho}+\bar{q}(\bar{\mathcal{L}}-\bar{\mathcal{L}'})e^{-\rho})(e^{\rho}+qe^{-\rho}(\mathcal{L}-\mathcal{L}'))\right.\nn\\
&&\left.+(e^{-\rho}(\mathcal{L}-\mathcal{L}')+\bar{q}e^{\rho})(-qe^{\rho}-e^{-\rho}(\bar{\mathcal{L}}-\bar{\mathcal{L}'}))\right\}dx^+dx^-+d\rho^2\label{me2}
\eea
The equations of motion from the action (\ref{action3}) are just
\be
R^{(i)}_{\mu\nu}=-2g^{(i)}_{\mu\nu}, \hs{3ex}i=1,2. \label{eom2}
\ee
One can check that the solution (\ref{me1},\ref{me2}) indeed satisfy the equation (\ref{eom2}).
To see more clearly what kinds of solutions (\ref{me1}) and (\ref{me2}) are, it is
 convenient to redefine the coordinates
\be
\omega=x^++qx^-,\ \bar{\omega}=x^-+\bar{q}x^+,\ \omega'=x^+-qx^-,\ \bar{\omega}'=x^--\bar{q}x^+\label{rep}
\ee
then the solutions (\ref{me1}) and (\ref{me2}) can be recast into
\bea
ds_1^2&=&d\rho^2+(\mathcal{L}+\mathcal{L}')d\omega^2+(\bar{\mathcal{L}}+\bar{\mathcal{L}'})d\bar{\omega}^2-(e^{2\rho}+(\mathcal{L}+\mathcal{L}')(\bar{\mathcal{L}}+\bar{\mathcal{L}'})e^{-2\rho})d\omega d\bar{\omega},\label{me1'}\\
ds_2^2&=&d\rho^2+(\mathcal{L}-\mathcal{L}')d\omega'^2+(\bar{\mathcal{L}}-\bar{\mathcal{L}'})d\bar{\omega'}^2-(e^{2\rho}+(\mathcal{L}-\mathcal{L}')(\bar{\mathcal{L}}-\bar{\mathcal{L}'})e^{-2\rho})d\omega'd\bar{\omega'}.\nn\\
\label{me2'}
\eea
These are exactly the BTZ black holes, the solutions of the AdS$_3$ gravities (\ref{eom2}). Note that the above coordinate transformations are allowed as we have separated diffeomorphism on $g^{(1)}_{\mu\nu}$ and $g^{(2)}_{\mu\nu}$. 
As we have discussed before, especially from the first order formalism, the suitable interpretation of this solution is that it is a black hole with a nonzero spin-2 charge. The metric and the spin-2 field are related to the above solutions by
\bea
ds^2&=&\frac{1}{2}(ds^2_1+ds^2_2)\nn\\
&=&d\rho^2+(\mathcal{L}+\bar{q}^2\bar{\mathcal{L}}-\bar{q}(\mathcal{L}\bar{\mathcal{L}}'+\mathcal{L}'\bar{\mathcal{L}})e^{-2\rho})(dx^+)^2\nn\\
&&+(\bar{\mathcal{L}}+q^2\mathcal{L}-q(\mathcal{L}\bar{\mathcal{L}}'+\mathcal{L}'\bar{\mathcal{L}})e^{-2\rho})(dx^-)^2\label{metric}\\
&&+(-(1+q\bar{q})(e^{2\rho}+(\mathcal{L}\bar{\mathcal{L}}+\mathcal{L}'\bar{\mathcal{L}}')e^{-2\rho})+2q\mathcal{L}'+2\bar{q}\bar{\mathcal{L}}')dx^+dx^-\nn\\
ds'^2&=&\frac{1}{2}(ds^2_1-ds^2_2)\nn\\
&=&(\mathcal{L}'+\bar{q}^2\bar{\mathcal{L}}'-\bar{q}(e^{2\rho}+(\mathcal{L}\bar{\mathcal{L}}+\mathcal{L}'\bar{\mathcal{L}}')e^{-2\rho}))(dx^+)^2\nn\\
&&+(\bar{\mathcal{L}}'+q^2\mathcal{L}'-q(e^{2\rho}+(\mathcal{L}\bar{\mathcal{L}}+\mathcal{L}'\bar{\mathcal{L}}')e^{-2\rho}))(dx^-)^2\label{spin2}\\
&&+(2q\mathcal{L}+2\bar{q}\bar{\mathcal{L}}-(1+q\bar{q})(\mathcal{L}\bar{\mathcal{L}}'+\mathcal{L}'\bar{\mathcal{L}})e^{-2\rho})dx^+dx^-\nn
\eea
The immediate question is how to analyze this configuration in the second order formalism. This will be the content of the next subsection.

\subsection{Black holes}

In the previous subsection, we have shown that the action (\ref{action3}) is well defined and it describes a system of a spin-2 field coupling to the gravity. Though effectively, this theory can be described as two decoupled fields $g^{(1)}$ and $g^{(2)}$, only their combination $\frac{1}{2}(g^{(1)}+g^{(2)})$ can be rightly interpreted as a spacetime metric. Therefore,  the system describes a complicated interaction between the graviton $g_{\mu\nu}$ and the spin-2 field $h_{\mu\nu}$. From the lower spin point of view, it can be regarded as the extension of the theory that describes a scalar or a Maxwell field coupling to gravity. However, for the theory with a scalar or Maxwell field coupled to gravity, the gauge transformation of the lower spin is independent of the diffeomorphism. But for the theory we consider, the spin-2 gauge transformation mixes with the diffeomorphism. So from this point of view, our theory should be considered as the simplest theory of the higher spin gravity.

 Usually the interacting higher spin theory is best developed in the first order formalism, while the possible second order formalism is quite involved. Our theory presents the first example of a ``high" spin gravity with a well-defined second order formulation. It would be interesting to discuss the black hole physics from the point of view of the metric and the spin-2 fields, even though the conventional notions of spacetime could be modified.


The solutions (\ref{me1},\ref{me2}) or equivalently (\ref{metric},\ref{spin2}) involve six parameters and hence are not easy to deal with. To simplify the problem, we first consider the nonrotating case, which corresponds to \footnote{Here nonrotating just means $g_{t\phi}=0$. There are other cases that can also lead to $g_{t\phi}=0$, but we ignore those cases.}
\be
\mathcal{L}=\bar{\mathcal{L}},\ \mathcal{L}'=-\bar{\mathcal{L}'},\ q=-\bar{q},
\ee
which reduces the number of the parameters to three. Now the solutions become
\bea
ds^2_1&=&(e^{\rho}+q(\mathcal{L}-\mathcal{L}')e^{-\rho})(q e^{\rho}+(\mathcal{L}+\mathcal{L}')e^{-\rho})(dx^+)^2\nn\\
&&-(q e^{\rho}-(\mathcal{L}-\mathcal{L}')e^{-\rho})(e^{\rho}-q(\mathcal{L}+\mathcal{L}')e^{-\rho})(dx^-)^2\nn\\
&&+\left\{-(e^{\rho}+q(\mathcal{L}-\mathcal{L}')e^{-\rho})(e^{\rho}-q(\mathcal{L}+\mathcal{L}')e^{-\rho})\right.\nn\\
&&\left.+(q e^{\rho}+(\mathcal{L}+\mathcal{L}')e^{-\rho})(q e^{\rho}-(\mathcal{L}-\mathcal{L}')e^{-\rho})\right\}dx^+dx^-+d\rho^2,\\
ds^2_2&=&(e^{\rho}-q(\mathcal{L}+\mathcal{L}')e^{-\rho})(e^{-\rho}(\mathcal{L}-\mathcal{L}')-qe^{\rho})(dx^+)^2\nn\\
&&+(qe^{\rho}+(\mathcal{L}+\mathcal{L}')e^{-\rho})(e^{\rho}+q(\mathcal{L}-\mathcal{L}')e^{-\rho})(dx^-)^2\nn\\
&&+\left\{-(e^{\rho}+q(\mathcal{L}-\mathcal{L}')e^{-\rho})(e^{\rho}-q(\mathcal{L}+\mathcal{L}')e^{-\rho})\right.\nn\\
&&\left.+(q e^{\rho}+(\mathcal{L}+\mathcal{L}')e^{-\rho})(q e^{\rho}-(\mathcal{L}-\mathcal{L}')e^{-\rho})\right\}dx^+dx^-
+d\rho^2.
\eea
Going to the physical fields $g_{\mu\nu}$ and $h_{\mu\nu}$, we find
\bea
ds^2&=&d\rho^2+(1+q^2)\mathcal{L}[(dx^+)^2+(dx^-)^2]\nn\\
&&+[-(1-q^2)(e^{2\rho}+(\mathcal{L}+\mathcal{L}')(\mathcal{L}-\mathcal{L}')e^{-2\rho})+4q\mathcal{L}']dx^+dx^-,\\
ds'^2&=&q(e^{2\rho}+(\mathcal{L}+\mathcal{L}')(\mathcal{L}-\mathcal{L}')e^{-2\rho})[(dx^+)^2-(dx^-)^2].
\eea
This configuration is like a RN black hole, but now the charge is from a spin-2 gauge field rather than the spin-1 gauge field.

It is not {\it a priori} evident that we can determine all the physical quantities from the metric and the spin-2 field as  in conventional Einstein gravity due to the mixing of diffeomorphism and the spin-2 gauge transformation. However, we can try to do this and to see whether we can get the same result as the one in the first order formulation.

The first quantity we need to know is the temperature. Let us first use $x^{\pm}=t\pm\phi$ to find the nonzero component of the metric
\bea
g_{tt}&=&2(q^2+1)\mathcal{L}+[-(1-q^2)(e^{2\rho}+(\mathcal{L}+\mathcal{L}')(\mathcal{L}-\mathcal{L}')e^{-2\rho})+4q\mathcal{L}'],\\
g_{\phi\phi}&=&2(q^2+1)\mathcal{L}-[-(1-q^2)(e^{2\rho}+(\mathcal{L}+\mathcal{L}')(\mathcal{L}-\mathcal{L}')e^{-2\rho})+4q\mathcal{L}'],\\
g_{\rho\rho}&=&1
\eea
and the nonzero spin-2 field is
\be
h_{t\phi}=h_{\phi t}=2q(e^{2\rho}+(\mathcal{L}+\mathcal{L}')(\mathcal{L}-\mathcal{L}')e^{-2\rho}).
\ee
We denote the horizon at $\rho=\rho_+$. In the conventional way to determine the temperature,  we require
\be
g_{tt}(\rho_+)=0,\ \partial_{\rho}g_{tt}(\rho_+)=0,
\ee
which determine the horizon $\rho_+$ and the parameter $q$ in terms of $\mathcal{L}$ and $\mathcal{L}'$
\bea
e^{2\rho_+}=\sqrt{(\mathcal{L}+\mathcal{L}')(\mathcal{L}-\mathcal{L}')},\\
q=\frac{\sqrt{\mathcal{L}-\mathcal{L}'}-\sqrt{\mathcal{L}+\mathcal{L}'}}{\sqrt{\mathcal{L}-\mathcal{L}'}+\sqrt{\mathcal{L}+\mathcal{L}'}}.\label{q2}
\eea
Note that the relation (\ref{q2}) shows that $q$ is not an independent parameter, as we have known from first order formulation. More interestingly, it is exactly the one (\ref{q2}) found in the first order formulation
in the non-rotating case. This motivates us to go further to determine the temperature $T=\frac{1}{\beta}$. We define $r=\rho-\rho_+$ and expand the metric around the horizon and go to the Euclidean time $t\to i t_E$, then we find
\be
ds^2\sim dr^2-\frac{1}{2}\partial_{\rho}^2g_{tt}(\rho_+)r^2dt_{E}^2+g_{\phi\phi}(\rho_+)d\phi^2
\ee
The regularity at the horizon gives us the temperature 
\be
\beta=2\pi\sqrt{\frac{2}{-\partial_{\rho}^2g_{tt}(\rho_+)}}=\frac{\pi}{2}(\frac{1}{\sqrt{\mathcal{L}+\mathcal{L}'}}+\frac{1}{\sqrt{\mathcal{L}-\mathcal{L}'}}),\label{beta}
\ee
which is consistent with the
 result in the first order formulation by using
\be
\beta=-i2\pi\tau=-i2\pi\frac{2\pi}{k}\frac{i}{4\pi^2}\frac{\partial S}{\partial \mathcal{L}}=\frac{\pi}{2}(\frac{1}{\sqrt{\mathcal{L}+\mathcal{L}'}}+\frac{1}{\sqrt{\mathcal{L}-\mathcal{L}'}}).
\ee

The next question is whether we can determine the entropy from the horizon area.
The horizon area and hence the corresponding entropy $S'$ is
\be
S'=\frac{Area}{4G}=\frac{1}{4G}\int_0^{2\pi}d\phi \sqrt{g_{\phi\phi}(\rho_+)}=\frac{8\pi k\mathcal{L}}{\sqrt{\mathcal{L}+\mathcal{L}'}+\sqrt{\mathcal{L}-\mathcal{L}'}}.
\ee
On the other hand the entropy in the non-rotating case from the first order formulation gives us
\be
S=2\pi k(\sqrt{\mathcal{L}+\mathcal{L}'}+\sqrt{\mathcal{L}-\mathcal{L}'}).
\ee
The two entropies are in match only when $\mathcal{L}'=0$ and for general $\mathcal{L}'$, we always have
\be
S<S'\label{entropy2}.
\ee
When $\mathcal{L}'=0$, the chemical potential $q$ and the spin-2 field vanishes, then the configuration becomes a pure black hole.

As the conventional notions of general relativity become impotent due to the mixing of the diffeomorphism with the spin-2 gauge transformations, we may try some other ways to  explore the thermodynamics of the black hole. We would like to use the Euclidean path integral which was developed in \cite{Harking:1983} to study this issue.  The partition function is of the form
\be
Z \sim \int [{\cal D}g_{\mu\nu}][{\cal D}h_{\mu\nu}]\exp(I_E).
\ee
Formally we need to do a path integral over $g_{\mu\nu}$ and $h_{\mu\nu}$. Due to the involved form of the action in terms of $g_{\mu\nu}$ and $h_{\mu\nu}$, this seems to be very difficult. Fortunately, here we are only interested in the tree-level contribution so we just need to consider the saddle point approximation.
By evaluating the Euclidean action and subtracting the counterterm coming from the AdS$_3$ vacuum, we can give the partition function $Z$ as
\be
Z\sim \exp(I_E^r)
\ee
where $I^r_E$ is the infinite volume limit of the regularized action
\be
I(R)=[-\frac{1}{8\pi G}\int_{0}^{\beta}dt_{E}\int_{0}^{2\pi}\int_{\rho_{+}}^{R}(\sqrt{g^{(1)}}+\sqrt{g^{(2)}})d\rho]-[-\frac{1}{8\pi G}\int_0^{\beta_r}dt_{E}\int_{0}^{2\pi}\int_{-\infty}^{R}(2\sqrt{g_{AdS}})d\rho]\label{Inter}
\ee
where we have introduced a cutoff $R$ and a regularization term coming from thermal AdS$_3$ spacetime. Note that in principle there is a surface term \cite{York:1972,Harking:1977} to make the theory  have a well-defined variation. This surface term is important in the asymptotic flat space\cite{Harking:1977} and contributes the whole path integral classically for the Schwarzschild black hole. This surface term is not important in AdS$_3$ as it has no contribution classically. However, in (\ref{Inter}) there is another tricky point. In the integral we have introduced an upper bound $\beta_r$ in the integral over $t_E$. This bound can be determined by the requirement that the line element at the surface $\rho=R$ should be the same for the configuration and the vacuum. If we used an ``averaged" metric $\frac{1}{2}(\sqrt{g^{(1)}}+\sqrt{g^{(2)}})$ to compute the line element, we require that
\be
\beta (\sqrt{g^{(1)}}+\sqrt{g^{(2)}})|_{\rho=R}=\beta_r (2\sqrt{g_{AdS}})|_{\rho=R}, \label{bound1}
\ee
then we find that $\beta_r=\beta(1+\mathcal{O}(e^{-4R}))$. Such an ``averaged" metric is expected to encode the information of both the spacetime and the spin-2 field. In the above treatment, the underlying guideline we follow is that the spacetime is intact. Actually there is another choice on the bound $\beta_r$. As  the spacetime background is now of metric $g_{\mu\nu}$, it seems natural to use
\be
\beta \sqrt{g}|_{\rho=R}=\beta_r (\sqrt{g_{AdS}})|_{\rho=R}. \label{bound2}
\ee
This gives us once again  $\beta_r=\beta(1+\mathcal{O}(e^{-4R}))$. Therefore the two choices make no difference
on this point.

 Taking all the facts above into account, doing the integral (\ref{Inter}) and taking the large $R$ limit, we then obtain
\be
I^r_E=\frac{4\pi k\mathcal{L}}{\sqrt{\mathcal{L}+\mathcal{L}'}+\sqrt{\mathcal{L}-\mathcal{L}'}},
\ee
and the corresponding partition function
\be
\ln Z'=I^r_E=\frac{4\pi k\mathcal{L}}{\sqrt{\mathcal{L}+\mathcal{L}'}+\sqrt{\mathcal{L}-\mathcal{L}'}}.
\ee
On the other hand,  from the first order formulation we have  the partition function
\be
\ln Z=\pi k(\sqrt{\mathcal{L}+\mathcal{L}'}+\sqrt{\mathcal{L}-\mathcal{L}'}).
\ee
The two partition functions are only in match when $\mathcal{L}'=0$ which means zero spin-2 charge. In that case, the theory becomes pure gravity and the results from the two formulations could be the same. For arbitrary other $\mathcal{L}'$, the two partition functions are different
 \be
\ln Z<\ln Z' .
\ee
Note that the discrepancy is quite similar to the discrepancy on the entropies in (\ref{entropy2}). These two kinds of discrepancies imply that the naively using conventional concepts and techniques in the Einstein gravity may lead to a wrong result. This is actually what we expected, since there is no simple geometrical interpretation now.  In retrospect, this geometric guideline is reasonable but not justified. In particular, when we consider more general solutions, the geometric picture may break down. For example,  the spacetime metric (\ref{metric}) generically has no horizon in original coordinates. One needs to take some gauge transformations to make a horizon, similar to the spin-3 black hole. In these general cases, the above geometric treatment is problematic.

The problem comes from the fact that if we interpret the theory as a spin 2 matter $h_{\mu\nu}$ coupling to $g_{\mu\nu}$, then there is no simple geometric picture. However, an alert reader may notice that the action (\ref{action3}) is actually the one that appeared in the decoupled bi-gravity\cite{salam}. The theory with multiple massless gravitons has been ignored partially due to the fact that 
  the two metric fields are independent of each other. However, we revisit this theory below.

Through the field redefinition (\ref{redef}) the partition function could be cast into the form
\bea
Z &\sim& \int [{\cal D}g^{(1)}_{\mu\nu}][{\cal D}g^{(2)}_{\mu\nu}]\exp(I)\nn\\
 &=&\int [{\cal D}g^{(1)}_{\mu\nu}] \exp(I^{(1)}) \int [{\cal D}g^{(2)}_{\mu\nu}] \exp(I^{(2)}).
\eea
where we have used the fact that the two fields $g^{(i)}_{\mu\nu}$ are completely decoupled and the action could be decomposed into two parts with
\be
I^{(i)}=\frac{1}{32\pi G} \int d^3x \sqrt{-g^{(i)}}(R^{(i)}+2), \ i=1,2.
\ee
Now the partition function is just the product of the partition functions of two decoupled systems
\be
Z=Z^{(1)}\times Z^{(2)} \label{part2}
\ee
with
\be
Z^{(i)}=\int [{\cal D}g^{(i)}_{\mu\nu}] \exp(I^{(i)}).
\ee
As the transformation (\ref{redef}) is completely linear, we do not worry that it may bring troubles. There is great advantage to working with the partition function (\ref{part2}) as it reduces to the usual one in Euclidean quantum gravity, with the standard gauge transformations on the metric fields $g^{(i)}_{\mu\nu}$. There seem to be two parallel worlds, each with its own metric and diffeomorphism. Hence, we expect that the geometric notions can be used in each world. That is to say, the observer in each world will use his/her own metric to talk about what he/she sees. There is no guarantee that they will use the same coordinates $t,\phi$ as we used before. Let us consider the solutions (\ref{me1}) and (\ref{me2}). The physical meaning of these two solutions is more transparent in the forms (\ref{me1'}) and (\ref{me2'}), which show that they are actually BTZ black holes. More explicitly, the  observer in the first world establishes his/her own coordinate and he/she finds the first BTZ black hole while the second observer finds another BTZ black hole. However, we should be more careful about this point. If we use Lorentz signature naively, namely, set $\omega=t'+\phi',\bar{\omega}=t'-\phi',\omega'=t''+\phi'', \bar{\omega}'=t''-\phi''$ and use (\ref{rep}), we will find a puzzle
 \bea
 t'=t+\frac{1}{2}(q+\bar{q})t+\frac{1}{2}(\bar{q}-q)\phi,\ \phi'=\phi-\frac{1}{2}(q+\bar{q})\phi+\frac{1}{2}(q-\bar{q})t,\\
 t''=t-\frac{1}{2}(q+\bar{q})t+\frac{1}{2}(q-\bar{q})\phi,\ \phi''=\phi+\frac{1}{2}(q+\bar{q})\phi+\frac{1}{2}(\bar{q}-q)t.
 \eea
 It seems that we cannot take $\phi',\phi''$ as angular variables since they contain $t$ which is noncompact. Actually this just shows that $t, \phi$ are ill defined in each world. An observer in the first world never cares about the relation between $\omega,\bar{\omega}$ and $t,\phi$, he/she defines his/her own coordinates $t',\phi'$ from $\omega,\bar{\omega}$. The same goes for the observer in the second world. That is to say,
     going to the Euclidean signature $x^+\to z, x^-\to -\bar{z}$, we have
\bea
\omega\to z_1=z-q\bar{z},\ & & \bar{\omega}\to-\bar{z}_1=-\bar{z}+\bar{q}z,\ \nn\\
\omega'\to z_2=z+q\bar{z},\ & & \bar{\omega}'\to-\bar{z}_2=-\bar{z}-\bar{q}z
\eea
Using the relation of $q,\bar{q}$ and the chemical potential $\mu',\bar{\mu}'$ which can be determined by $\tau,\bar{\tau},\alpha,\bar{\alpha}$, we find the identification
\be
z_1\sim z_1+2\pi(\tau+\alpha),\ \bar{z}_1\sim \bar{z}_1+2\pi(\bar{\tau}+\bar{\alpha}),\ z_2\sim z_2+2\pi(\tau-\alpha),\ \bar{z}_2\sim\bar{z}_2+2\pi(\bar{\tau}-\bar{\alpha})
\ee
With these identifications,  the form of the black hole solution given in (\ref{me1'},\ref{me2'}) are exactly the Euclidean BTZ black holes. These solutions are non-singular 
 when $z_1,\bar{z}_1$ are related to an angle $\phi'$ and Euclidean time $t_{E}'$ while $z_2,\bar{z}_2$ are related to another angle $\phi''$ and Euclidean time $t_{E}''$ with the identification
\bea
\phi'\sim\phi'+2\pi,\ t_{E}'\sim t_{E}'+\beta^{(1)},\\
\phi''\sim\phi''+2\pi,\ t_{E}''\sim t_{E}''+\beta^{(2)}
\eea
where $\beta^{(1)},\beta^{(2)}$ are determined by
\be
\beta^{(1)}=-i \pi[(\tau+\alpha)-(\bar{\tau}+\bar{\alpha})],\ \beta^{(2)}=-i\pi[(\tau-\alpha)-(\bar{\tau}-\bar{\alpha})]
\ee
In the non-rotating case they are
\be
\beta^{(1)}=\beta^{(2)}=\frac{\pi}{2}(\frac{1}{\sqrt{\mathcal{L}+\mathcal{L}'}}+\frac{1}{\sqrt{\mathcal{L}-\mathcal{L}'}}).
\ee
Then we use the Euclidean path integral to evaluate
\bea
I(R)&=&-\frac{1}{8\pi G}[\int_{0}^{\beta^{(1)}}dt'_E\int_{0}^{2\pi}d\phi'\int_{\rho_+}^{R}\sqrt{g^{(1)}}d\rho-\int_0^{\beta^{(1)}_r}dt'_E\int_{0}^{2\pi}d\phi'\int_{\rho_+}^{R}\sqrt{g_{AdS}}d\rho]+[1\to 2].\nn\\
\label{I1}
\eea
This time $\beta^{(1)}_r=\beta^{(1)}(1+\mathcal{O}(e^{-4R}))$. Taking the large $R$ limit we find
\be
\ln Z=I=\lim_{R\to\infty}I(R)=\pi k(\sqrt{\mathcal{L}+\mathcal{L}'}+\sqrt{\mathcal{L}-\mathcal{L}'}).
\ee
We find that the total partition function is consistent with the one found in the first order formulation.

There are several remarks in order:
\begin{enumerate}
\item Note that though the relation (\ref{I1}) looks similar to the relation (\ref{Inter}) with the same $\beta$, the metric has been modified, leading to a different result.
\item When working on the contributions from the $g^{(1)}$ and $g^{(2)}$ fields, we have to use  different sets of coordinates $z_1, \bar{z}_1$ and $z_2, \bar{z}_2$. This sounds strange but makes sense if we take the metrics $g^{(1)}$ and $g^{(2)}$ as independent fields in doing path-integral. Actually in this case, the choice of the coordinates is not important, as we are integrated over the fields $g^{(i)}_{\mu\nu}$ which have their own diffeomorphism.
\item Even though our discussion on the partition function was focused on the non-rotating black hole, the generic case could be studied similarly. The partition functions of $g^{(1)}$ and $g^{(2)}$ are just the ones of usual BTZ black holes. Thus their sum reproduces exactly the partition function (\ref{partion1}) in the first order formulation.
\item Physically, the picture is as follows. From the first order formulation, we found the black hole solution (\ref{BH}) with spin-2 hair. This black hole has the entropy (\ref{entropy'}), which is reminiscent of the entropies of two BTZ black holes. Note that in getting the  entropy (\ref{entropy'}), we have used the techniques in the high spin gravity, with the help of dual holographic OPE. On the other hand, the solution (\ref{BH}) corresponds to the BTZ black holes (\ref{me1}) and (\ref{me2}) in AdS$_3$ bi-gravity. These two black holes come from the excitations of two different metric fields, which are decoupled completely, so it is safe to treat every black hole independently. The partition function of these solutions in the bi-gravity theory is just the product of two independent partition functions, and the entropies of the solutions are just the sum of the ones of two independent BTZ black holes.
\end{enumerate}




\section{Dual description}

In this section, we consider the problem of the dual description of our theory in the second order formulation. In the first order formulation, the dual description has been investigated in Section 2.  The information we got is that there is a dual CFT with the central charge $6k$ and the asymptotical symmetry given in (\ref{OPE}). In this CFT, there is a stress tensor $\mathcal{T}$ and an additional spin-2 charge $\tilde{\mathcal{T}}$, as has been shown from the symmetry analysis with the result encoded in (\ref{OPE}). However, this statement is quite unusual since we have two spin-2 operators. A closer look at the result (\ref{OPE}) suggests that we redefine  two other spin-2 operators by $T_1=\frac{\mathcal{T}+\tilde{\mathcal{T}}}{2},\ T_2=\frac{\mathcal{T}-\tilde{\mathcal{T}}}{2}$, then we find that
\bea
T_1(z)T_1(0)&\sim&\frac{c/4}{z^4}+\frac{2T_1}{z^2}+\frac{\partial T_1}{z},\nn\\
T_1(z)T_2(0)&\sim&regular,\\
T_2(z)T_2(0)&\sim&\frac{c/4}{z^4}+\frac{2T_2}{z^2}+\frac{\partial T_2}{z}.\nn
\eea
From these OPE relations, we learn that there are two decoupled CFTs at the asymptotical boundary. These two CFTs have the same central charge $3k$, and there is no interaction between them.
For the bar term, we have similar results.

 It is illuminating to study  the same question in the second order formulation. Let us give a heuristic derivation following the reference $\cite{HSBH}$, or the lecture given in \cite{Kraus:2006}. Inspired by the action (\ref{action3}), we define two kinds of spin-2 operators by explicitly considering the surface term and doing subtraction as shown in \cite{Kraus:1999, Haro:2000}
 \be
 T^{(i)}_{\mu\nu}=\frac{1}{16\pi G}(\Theta^{(i)}_{\mu\nu}-\Theta^{(i)}\gamma^{(i)}_{\mu\nu}-\gamma^{(i)}_{\mu\nu}).
 \ee
The index $i$ corresponds to the field $g^{(i)}$. The $\gamma^{(i)}$'s are the boundary fields induced by $g^{(i)}$. $\Theta^{(i)}_{\mu\nu}$ is the extrinsic curvature constructed by $g^{(i)}$ and $\Theta$ is its trace evaluated by $\gamma^{(i)}$.  The next step is to analyze the asymptotic symmetry. Note that the allowed symmetry transformation has been identified in the previous subsection and asymptotic symmetry analysis can be done conveniently in the coordinate (\ref{rep}). The discussion is straightforward, as shown originally by
 Brown-Henneaux  \cite{Brown:1986}. The result is that under the symmetry transformation,
\be
T_{\omega\omega}\to T_{\omega\omega}+2\partial_{\omega}\epsilon(\omega)T_{\omega\omega}+\epsilon(\omega)\partial_{\omega}T_{\omega\omega}-\frac{c}{24\pi}\partial^3_{\omega}\epsilon(\omega)
\ee
with $c=\frac{3}{4G}=3k$ due to the rescaling the the coefficients. We have omitted the index $i$ as it is true for both $g^{(i)}_{\mu\nu}$. There is a similar result for the right-moving sector. Note that this is the correct Virasoro transformation of a spin-2 operator. Therefore we reach a consistent picture with the first order formulation.

Even though the decoupled CFTs could be obtained by redefining the operators of the original CFT, the interpretation is very different. For the $so(2,2)$ Chern-Simons gravity, the dual CFT is of central charge $6k$ and has an extra spin-2 operator. From the boundary CFT point of view, the black hole in this theory could be understood as the excited CFT at the finite temperature ($1/\beta_L, 1/\beta_R$) and of the chemical potential $(\mu', \bar{\mu}')$. On the other hand, from the point of view of AdS$_3$ bi-gravity, there are two independent BTZ black holes, each could be described by a CFT.  The first CFT has a left-moving sector with central charge $3k$ and at level $\frac{k(\mathcal{L}+\mathcal{L}')}{2}$, and a right-moving sector with central charge $3k$ and at level $\frac{k(\bar{\mathcal{L}}+\bar{\mathcal{L}'})}{2}$. The second CFT has a left-moving sector with central charge $3k$ and at level $\frac{k(\mathcal{L}-\mathcal{L}')}{2}$, and a right-moving sector with central charge $3k$ and at level $\frac{k(\bar{\mathcal{L}}-\bar{\mathcal{L}'})}{2}$. Using the Cardy formula,
\be
S=2\pi\sqrt{\frac{c L}{6}},
\ee
 we find that the total entropy is
 \be
 S_{en}=\pi k (\sqrt{\mathcal{L}+\mathcal{L}'}+\sqrt{\mathcal{L}-\mathcal{L}'}+\sqrt{\bar{\mathcal{L}}+\bar{\mathcal{L}'}}+\sqrt{\bar{\mathcal{L}}-\bar{\mathcal{L}'}}),
 \ee
 exactly the same as the one (\ref{entropy'}) determined from the holonomy condition in the first order formulation.

\section{Conclusion and Discussion}

In this paper, motivated by the higher spin gravity and holography, we investigated a theory with two spin-2 fields. This theory can be constructed both in the first order formulation and in the second order formulation. From the point of view of Chern-Simons higher spin gravity, the theory can be regarded as a theory based on the gauge group $D_2$. We constructed the corresponding black holes and analyzed their thermodynamics using the standard method in the higher spin gravity.

In the second order formulation, the action of our theory could be recast into the one of an AdS$_3$ bi-gravity with two decoupled massless gravitons, after field redefinition. The black hole solutions found in the first order formulation could be conveniently interpreted as
two decoupled BTZ black holes in bi-gravity theory. The entropy and the asymptotical symmetry support this picture. Our study actually sheds some light on the decoupled bi-gravity theory. For this kind of theory, the
equations of motion for different metric fields are decoupled so there are various black hole solutions, which could be excitations of different metric fields. In such a system, the entropy should be just the sum of two black holes. Such a picture is most easily clear in the path integral formalism.
On the other hand, the successful reproduction of the entropy of the black hole with the spin-2 hair in the second order formulation strongly supports that the treatment of the higher spin black hole thermodynamics in terms of gauge invariant quantities is robust.

From the point of view of AdS/CFT, our investigation is related to another interesting problem. If we have a  CFT$_1$ at the boundary associated with a dual bulk description, and we also have a second CFT$_2$ at the boundary associated with another
 dual bulk description, let us put the two CFTs at the same boundary together. Note that this operation
 is well-defined. Then what is the dual bulk description? In the simplest case,
  we assume that two CFTs are decoupled, then the boundary theory has two kinds of stress tensors.
  From the general principle of AdS/CFT, there should be two kinds of spin-2 fields in the bulk, leading to a decoupled bi-gravity theory.
 We expect that this is a starting point to study more interesting case. One interesting case is  when there is some interaction between the CFTs; this may raise an interesting question about what the possible gravity theory could be.


 Our motivation to study this theory is partly based on the fact that it can be reformulated in a second order formulation.   The second order formulation of the higher spin gravity
is important for us because the physical meanings of fields are more transparent. And we expect to see the notions
 of spacetime geometry change explicitly in the second order formulation. Many aspects of higher spin gravity are indeed
 inherited in our model if we interpret the theory as a spin-2 matter coupling to the gravity.
 For the simplest case, we managed to rederive some results from
  the geometric notions. In the non-rotating black hole case, we found that we could use the
 regular horizon condition to determine the position of the horizon and read the value of the parameter $q$ and the black hole temperature. However for more generic cases, such treatments are useless, just as the case of  the higher spin black hole. Even for the simplest case, the area law of the black hole horizon does not give the correct black hole entropy, suggesting that the notion of horizon cannot be taken seriously.
In principle, from the second order formulation, it could be possible to read the global charges of the black holes, following the traditional way.  


\vspace*{10mm}
\noindent {\large{\bf Acknowledgments}}\\
The work was in part supported by NSFC Grants No. 10975005, and No. 11275010.
\vspace*{5mm}

\section*{Appendix: About Stokes Theorem}
The usual Stokes theorem states that
\be
\int_{\mathcal{M}}\sqrt{|g|}d^nx\ \nabla_{\mu}V^{\mu}=\int_{\partial\mathcal{M}}\sqrt{|\gamma|}d^{n-1}x \ n_{\mu}V^{\mu}
\ee
In our case, we need to show the the integral
\be
\int_{\mathcal{M}}\sqrt{-g^{(i)}}d^nx\ \nabla^{(i)}_{\mu}V^{\mu}\label{Integral}
\ee
still gives a boundary integral. We can show it by the following two formulas
\bea
\nabla^{(i)}_{\mu}V^{\mu}=\partial_{\mu}V^{\mu}+\Gamma^{(i)\mu}_{\ \ \mu\lambda}V^{\lambda}\\
\Gamma^{(i)\mu}_{\ \ \mu\lambda}=\frac{1}{\sqrt{-g^{(i)}}}\partial_{\lambda}\sqrt{-g^{(i)}}
\eea
Then the integral (\ref{Integral}) becomes
\be
\int_{\mathcal{M}}\partial_{\mu}(\sqrt{-g^{(i)}}V^{\mu}),
\ee
 a boundary term after integration.


\begin{thebibliography}{}



 \bibitem{Fronsdal1:1978}
 C.Fronsdal,
 ``Massless fields with interger spin'', Phys. Rev. D18(1978),3624-3629.

 \bibitem{Fronsdal2:1978}
 J.Fang and C.Fronsdal,
 ``Massless fields with half interger spin'', Phys. Rev. D18(1978),3630-3633.

 \bibitem{Bouatta:2004}
 N.Bouatta, G.Compere and A.Sagnotti,
 ``An introduction to freee higher spin fields'', hep-th/0409068.

\bibitem{Vasiliev}
 X. Bekaert, S. Cnockaert, C. Iazeolla and M. A. Vasiliev,
 ``Nonlinear higher spin theories in various dimensions ''arXiv:0503128.

\bibitem{ASG II:2010}
A. Campoleoni, S. Fredenhagen, S. Pfenninger and S. Theisen, ``Asymptotic sym-
metries of three-dimensional gravity coupled to higher-spin fields,'' JHEP 1011, 007
(2010) [arXiv:1008.4744 [hep-th]].

\bibitem{Blencowe:1989}
M. P. Blencowe, ``A Consistent Interacting Massless Higher Spin Field Theory In D
= (2+1),'' Class. Quant. Grav. 6, 443 (1989).
\bibitem{Blencowe:1990}
 E. Bergshoeff, M. P. Blencowe and K. S. Stelle, ``Area Preserving Diffeomorphisms
And Higher Spin Algebra,'' Commun. Math. Phys. 128, 213 (1990).

\bibitem{truncated}
Bin Chen, Jiang Long and Yi-Nan Wang,
``Black Holes in Truncated Higher Spin AdS$_3$ Gravity'',arXiv:1209.6185.

\bibitem{ASG I:2010}
M. Henneaux and S.-J. Rey, ``Nonlinear Winfinity as Asymptotic Symmetry of
Three-Dimensional Higher Spin Anti-de Sitter Gravity,'' JHEP 1012 (2010) 007,
arXiv:1008.4579.





\bibitem{Per kraus:2011}
  M.~Gutperle and P.~Kraus,
  ``Higher Spin Black Holes,''
  JHEP {\bf 1105}, 022 (2011).

 \bibitem{Ammon:2011nk}
  M.~Ammon, M.~Gutperle, P.~Kraus and E.~Perlmutter,
  ``Spacetime Geometry in Higher Spin Gravity,''
  arXiv:1106.4788.

\bibitem{Campoleoni:2012hp}
  A.~Campoleoni, S.~Fredenhagen, S.~Pfenninger and S.~Theisen,
  ``Towards metric-like higher-spin gauge theories in three dimensions,''  arXiv:1208.1851 [hep-th].

\bibitem{Ippei Fujisawa}
Ippei Fujisawa and Ryuichi Nakayama,
``Second-Order Formalism for 3D Spin-3 Gravity'',arXiv:1209.0894 [hep-th].



 \bibitem{Klevanov}
I. Klebanov and A. Polyakov,``$AdS$ dual of the critical O(N) vector model,''
Phys. Lett. B550(2002)213-219, arXiv:hep-th/0210114.

\bibitem{Giombi:2012he}
  S.~Giombi and X.~Yin,
  ``The Higher Spin/Vector Model Duality,''  arXiv:1208.4036 [hep-th].

  \bibitem{salam}
  C. J. Isham, A. Salam, and J. A. Strathdee,
  ¡°F-dominance of gravity,¡± Phys. Rev., D3:867, (1971).

\bibitem{Hassan:2011}
S.F. Hassan and Rachel A. Rosen,
``Bimetric Gravity from Ghost-free Massive Gravity'',  JHEP 1202 (2012) 126.

\bibitem{Hinterbichler:2012}
Kurt Hinterbichler and  Rachel A. Rosen,
``Interacting Spin-2 Fields'', JHEP 1207 (2012) 047.



\bibitem{Henneaux:2000}
N. Boulanger, T. Damour, L. Gualtieri, and M. Henneaux, ``Inconsistency of
interacting, multigraviton theories'', Nucl.Phys. B597 (2001) 127-171,
arXiv:hep-th/0007220 [hep-th].

\bibitem{Prokushkin:1998bq}
  S.~F.~Prokushkin and M.~A.~Vasiliev,
  ``Higher spin gauge interactions for massive matter fields in 3-D AdS space-time,''  Nucl.\ Phys.\ B {\bf 545}, 385 (1999)  [hep-th/9806236].

\bibitem{Chang:2012kt}
  C.~-M.~Chang, S.~Minwalla, T.~Sharma and X.~Yin,
  ``ABJ Triality: from Higher Spin Fields to Strings,''  arXiv:1207.4485 [hep-th].

 \bibitem{Hstmg}
Bin Chen and Jiang Long,
``High Spin Topologically Massive Gravity'', JHEP 1112 (2011) 114 ,arXiv:1110.5113 [hep-th].

\bibitem{Harking:1983}
S. W. Hawking and D. Page, ``Thermodynamics Of Black Holes In Anti-de Sitter
Space,'' Commun. Math. Phys. 87 (1983) 577.

\bibitem{York:1972}
J. W. York, ``Role Of Conformal Three-Geometry In The Dynamics Of Gravitation,''
Phys. Rev. Lett. 28 (1972) 1082.

\bibitem{Harking:1977}
G. W. Gibbons and S. W. Hawking, ``Action Integrals And Partition Functions In
Quantum Gravity,'' Phys. Rev. D15 (1977) 2752.

\bibitem{HSBH}
Martin Ammon, Michael Gutperle, Per Kraus and Eric Perlmutter,
``Black holes in three dimensional higher spin gravity: A review ''
arXiv:1208.5182.

\bibitem{Kraus:2006}
P. Kraus, ``Lectures on black holes and the AdS(3)/CFT(2) correspondence,'' Lect.
Notes Phys. 755, 193 (2008), hep-th/0609074.

\bibitem{Kraus:1999}
V. Balasubramanian and P. Kraus, ``A stress tensor for anti-de Sitter gravity,'' Com-
mun. Math. Phys. 208, 413 (1999), hep-th/9902121.

\bibitem{Haro:2000}
S. de Haro, S. N. Solodukhin and K. Skenderis, ¡°Holographic reconstruction of space-
time and renormalization in the AdS/CFT correspondence,¡± Commun. Math. Phys.
217, 595 (2001),hep-th/0002230.


\bibitem{Brown:1986}
J. D. Brown and M. Henneaux, ``Central Charges in the Canonical Realization of
Asymptotic Symmetries: An Example from Three-Dimensional Gravity,'' Commun.
Math. Phys. 104, 207 (1986).

\end{thebibliography}
\end{document}